# Star formation quenching in massive galaxies

Understanding how and why star formation turns off in massive galaxies is a major challenge for studies of galaxy evolution. Many theoretical explanations have been proposed, but a definitive consensus is yet to be reached.

**Allison Man and Sirio Belli**

Despite the success of the Lambda Cold Dark Matter (ΛCDM) cosmological model in reproducing the observable Universe, certain properties of galaxies remain unexplained in this framework. Specifically, in order to reproduce the observed population of massive galaxies, cosmological models must include a poorly constrained quenching mechanism — a process that suppresses star formation — to solve two problems. The first one is the discrepancy between the observed galaxy mass function and the theoretical halo mass function[1]. The second one is the observation that more massive galaxies have systematically older stars[2]. This contravenes the hierarchical nature of ΛCDM, in which more massive galaxies are expected to be younger, since they assemble at later times.

The term 'quenching' has been used with two different meanings in the literature, to indicate either the termination of the star formation activity, or the process of maintaining a galaxy quiescent over its lifetime, despite the fresh fuel produced by stellar evolution and gas inflows. Given the substantially different timescales involved in the two processes, it remains debated whether one single mechanism is responsible for both the onset and the maintenance of quenching. Archaeological evidence shows that the termination process must be particularly rapid for the most massive galaxies that would become present-day giant ellipticals, most of which were already quenched by z∼2.

In theoretical studies, feedback by energetic sources is often used to quench star formation. Though closely related, quenching and feedback are not strictly equivalent: there are theoretical scenarios in which galaxies are quenched by processes that are not feedback mechanisms, as we discuss below.

## Quenching mechanisms

Over the years, numerous mechanisms have been proposed to explain the quenching of massive galaxies, including active galactic nuclei (AGN), stellar feedback, and gravitational effects. Given that these processes encompass several branches of astrophysics, it is not surprising that studies on the topic have suffered from inconsistent definitions and confusion in the literature.

Here, we attempt to provide a concise overview and a comprehensive classification of the quenching mechanisms. Normally, stars form out of molecular gas of temperature $T < 10^2$ K, that cooled and settled from warm and hot gas within the halo ($T = 10^3 - 10^6$ K) previously accreted from cosmological filaments. Galaxy quenching can be understood as an interruption to any one of the necessary conditions to star formation, as illustrated in Fig. 1. Several of these processes may be in place simultaneously; it is the timescale that determines their relative importance[3].

Following the illustration in Fig. 1, we identify five broad classes of quenching mechanisms.

**(i) Gas does not accrete.** Massive galaxies may quench due to reduced gas accretion onto their dark matter haloes. This has been termed cosmological starvation[4], and stands as an example of quenching not driven by feedback. Even if accretion is completely shut off, the gas produced by stellar evolution could still fuel subsequent star formation. Therefore, an additional mechanism may be needed to maintain the galaxy quiescent.

**(ii) Gas does not cool.** In the standard picture of galaxy formation, gas collapses in a dark matter halo and heats up because of virial shocks[3]. However, simulations[5] show that the shocks are formed only when the halo is more massive than approximately $10^{12}$ solar masses. This mass scale is roughly the same above which galaxies are observed to be prevalently quiescent, suggesting that virial shock heating may account for the onset of star formation quenching.

After the shock, however, the hot halo gas should cool and become available for star formation. To prevent further star formation, additional gas heating is necessary. The most accepted explanation is feedback from radiatively inefficient accretion onto supermassive black holes (radio-mode feedback), that is only effective in massive halos with hot gas. Implementations of this feedback yielded the first successful mass function predictions from semi-analytical models[6,7]. In addition to black hole accretion, other gas heating mechanisms have been proposed. For example, a gas clump or a satellite falling into a massive halo releases potential energy. If this energy can be efficiently used to heat the halo gas via ram pressure or dynamical friction, this could constitute a substantial source of gravitational heating[8]. Evolved stellar populations may also contribute to the gas heating via stellar feedback

# What causes quenching in massive galaxies?

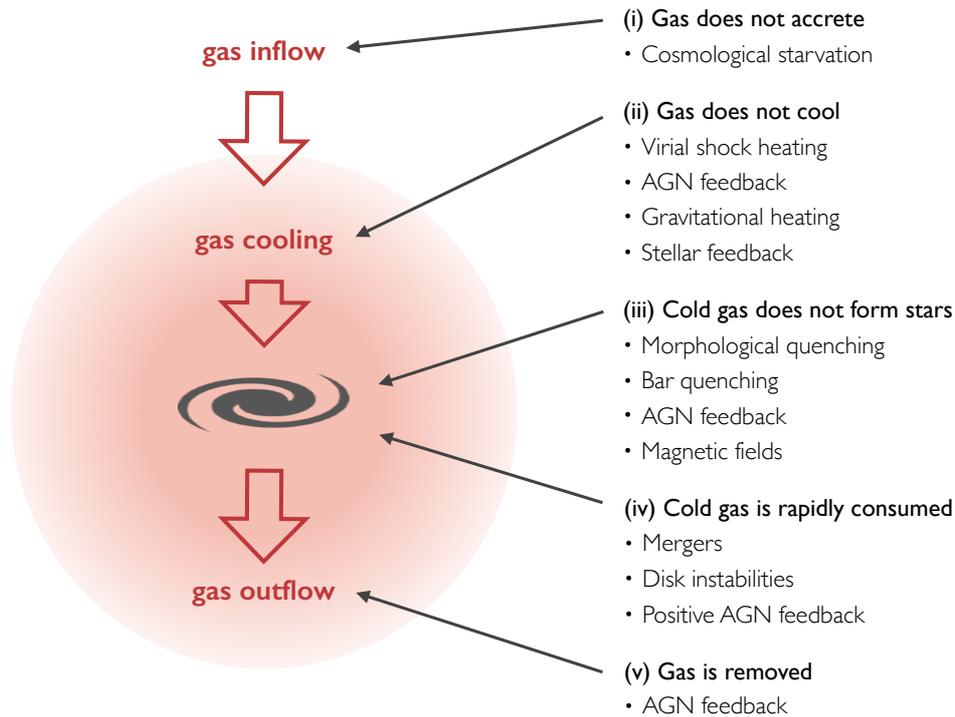

**Fig. 1 Schematic diagram listing the plausible quenching mechanisms.**

from Type Ia supernovae and asymptotic giant branch stars[9]. Regardless of the energy source, it is likely that only massive halos, in which the gas is roughly at the virial temperature, can be kept hot efficiently.

Heating processes are essential to solve a related problem. In galaxy clusters, the cooling time is short enough that the gas should quickly flow to the center and trigger intense star formation; but central galaxies are typically quiescent[10]. This 'cooling flow' problem is fundamentally similar to galaxy quenching but takes place on larger scales.

**(iii) Cold gas does not form stars efficiently.** Quenching may be due to a low star formation efficiency, rather than the absence of fuel. To form stars, cold gas must dissipate its kinetic energy (bulk motions and turbulence)[11]. Certain non-thermal processes may be capable of injecting significant amounts of kinetic energy, such that the dissipation timescale is too long for the gas to settle.

The stellar bulge could provide large-scale shear to inject turbulence, thereby stabilizing the gas disk from fragmentation. This process is called morphological quenching[12] (or sometimes 'gravitational quenching'), typically acts on timescales longer than 1 Gyr and is only effective at low molecular gas fractions, below about 10%. A similar mechanism can be caused by the stellar bar. This bar quenching may act on even shorter timescales if the stellar bar forms rapidly[13]. Other potential sources of turbulence injection are low-power AGN[14] and magnetic fields[15]. Such mechanisms may be responsible for maintaining low star formation rates, but are unlikely to be the sole explanation for the abrupt termination of star formation in massive galaxies.

**(iv) Cold gas is rapidly consumed.** If star formation consumes cold gas faster than it is replenished, the galaxy will run out of fuel and become quiescent. This can happen if star formation takes place in efficient bursts, potentially triggered by compressive gas motions and effective angular momentum loss. Major mergers of galaxies have first been invoked as a possible trigger[16], and more recently violent disk instabilities[17] and positive AGN feedback[18] have also been suggested. Clearly, gas consumption alone can quench star formation only temporarily. Additional mechanisms are required to maintain low star formation rates until the present day.

**(v) Gas is removed.** The accretion onto supermassive black holes could release sufficient energy and momentum to expel gas from galaxies. This quasar-mode feedback is most efficiently triggered by major mergers due to high gas inflow rates[19]. Although rare and short-lived, this process can quench star formation by removing the gas supply from massive galaxies.

We based our classification of quenching mechanisms on physical considerations, noting that each of the proposed mechanisms has a range of validity in terms of timescales, stellar masses, and gas masses. Observational limitations have also motivated a number of phenomenological classifications proposed in the literature. The most common one is the distinction between mass quenching, which affects massive galaxies, and environmental quenching, which only acts on satellite galaxies and is omitted from the present discussion[20]. Some studies instead use the term halo quenching to indicate a mechanism that relates to the halo mass rather than stellar mass. Another common division is that between internal and external quenching, but this division is ambiguously defined. Other studies distinguish between a slow quenching (also called strangulation or starvation), when some leftover cold gas is still available after the quenching event, and a fast quenching, which indicates an abrupt end of the star formation activity.

## Observational prospects

Semi-analytical models and, more recently, cosmological hydrodynamical simulations are capable of reproducing the basic properties of quiescent galaxies, such as number counts and colours. In all cases, this is made possible by some form of feedback, generally attributed to AGN, that heats up the gas only in halos above a mass threshold. Although this broad agreement is certainly encouraging, the fact that substantially different feedback implementations yield relatively similar galaxy populations suggests that the feedback recipes are currently degenerate[21]. AGN feedback could affect galaxy star formation through different physical processes (ii - v), and observational studies continue to challenge the validity of the crude feedback recipes applied in cosmological simulations[22]. The intricate connection between galaxies and AGN is crucial to understand quenching.

Observationally, the level of star formation suppression appears to correlate not only with stellar mass, but even more strongly with the surface mass density and central velocity dispersion. However, because galaxy properties are interrelated in a complex way, it is not straightforward to establish causal connections given the observed correlations.

As is clear from Fig. 1, studying the physical properties of the different gas phases represents a fundamental step towards understanding quenching. These challenging observations are particularly needed at high redshift, soon after the termination of star formation in massive galaxies. A pressing question is to determine whether recently quenched galaxies possess a significant amount of cold gas. If true, this would challenge the common assumption that galaxies stop forming stars because of a lack of gas, and would require explanations for the suppression of star formation efficiency. Another priority is to measure the properties of the hot gas in quiescent galaxies. If shock heating is prevalent in massive halos as predicted, one should expect to detect its signature in the circumgalactic medium. Spatially resolved observations could further constrain the origin of gas heating and shocks. Lastly, more detailed observations of gas outflows are needed to determine the importance of ejective processes. Outflows are known to be multi-phase, but it remains debated which phase carries the most mass, and whether most of the gas can escape the deep potential well of massive galaxies.

Valuable constraints on quenching can also be obtained with deep spectroscopic studies of the stellar content of quiescent galaxies. The star formation history contains a record on when and how quenching occurred. For example, observations of some nearby massive galaxies suggest episodes of rejuvenation, which would require a reversible quenching mechanism. Measurements of the stellar metallicity can also be used to probe the timescale of star formation and quenching. Overall, a thorough understanding of the star formation process is fundamental to solve the quenching puzzle.

Allison Man[1*] and Sirio Belli[2]

[1] *Dunlap Institute for Astronomy and Astrophysics, University of Toronto, Toronto, ON, Canada.*
[2] *Max-Planck-Institut für Extraterrestrische Physik, Garching bei München, Germany.*
*e-mail: [allison.man@dunlap.utoronto.ca](allison.man@dunlap.utoronto.ca)

## Acknowledgments

We thank the participants of the Lorentz Center workshop "The Physics of Quenching Massive Galaxies at High Redshift" held in Leiden, 6-10 November 2017, for insightful discussions that inspired this Comment. We are grateful to the Lorentz Center for hosting this workshop, and to M. Lehnert and T. Naab for commenting on this article.